\documentclass[%
    aip,
    jcp,
    amsmath,
    amssymb,
    twocolumn,
    superscriptaddress,
    reprint]{revtex4-1}
\usepackage{dcolumn}
\usepackage{graphicx} 
\usepackage{bm}
\usepackage{longtable}
\usepackage[english]{babel}
\usepackage{color}

\begin{document}

\title{Effective dielectric constant of water at the interface with charged C$_{60}$ fullerenes } 

\author{Setare Mostajabi Sarhangi} 
\affiliation{Molecular Simulation Research Laboratory, Department of Chemistry, Iran University of Science and Technology, Tehran 16846-13114, Iran}
\author{Morteza M.\ Waskasi} 
\affiliation{School of Molecular Sciences, Arizona State University, PO Box 871504, Tempe, AZ 85287-1504 }
\author{Seyed Majid Hashemianzadeh} 
\affiliation{Molecular Simulation Research Laboratory, Department of Chemistry, Iran University of Science and Technology, Tehran 16846-13114, Iran}
\author{Dmitry V.\ Matyushov }
\affiliation{Department of Physics and School of Molecular Sciences, Arizona State University,  PO Box 871504, Tempe, AZ 85287-1504 }
\email{dmitrym@asu.edu}

\begin{abstract}
Dipolar susceptibility of interfacial water and the corresponding interface dielectric constant were calculated from numerical molecular dynamics simulations for neutral and charged states of buckminsterfullerene C$_{60}$. Dielectric constants in the range 10--22, depending on temperature and solute charge, were found. The hydration water undergoes a structural crossover as a function of the solute charge. Its main signatures include the release of dangling O-H bonds pointing toward the solute and the change in the preferential orientations of hydration water from those characterizing hydrophobic to charged substrates. The interface dielectric constant marks the structural transition with a spike. The computational formalism adopted here provides direct access to interface susceptibility from configurations produced by computer simulations. The required property is the cross-correlation between the radial projection of the dipole moment of the solvation shell and the electrostatic potential of the solvent inside the solute.     
\end{abstract}
\maketitle

\section{Introduction}
\label{sec:1}
The dipolar susceptibility of a bulk material  $\chi$ is measured by the dielectric experiment in terms of the electrostatic free energy stored in a plane capacitor. The susceptibility defines the bulk dielectric constant\cite{Jackson:99} $\epsilon=1+4\pi\chi$.  Whether this material property can be applied to interfaces of molecular or mesoscopic dimensions has long been a subject of contention.\cite{Yeh:1999hb,Teschke:2005ea,VelascoVelez:2014ea,Shi:2016jq,Zhang:2018cf,Dreier:2018hp,Fumagalli:2018iw} It has long been suggested that an effective dielectric constant of a microscopic interface has to be introduced, and most researchers have agreed that this effective dielectric constant has to be reduced from the bulk value.\cite{Debye:1925dw,Conway:1951es,PALMER:1952ew,Stiles:1982ck,Ninham:1997gl,Brown:1999dx,PhysRevE.64.011605,Lenart:2007df,Wander:2008ck,Lima:2008dc,Boamah:2018kn} The extent of reduction has mostly remained unknown. Following earlier indications,\cite{PhysRevE.64.011605,Teschke:2005ea} direct measurements have been recently reported for the dielectric constant of water in contact with a graphite substrate as a function of the water film thickness.\cite{Fumagalli:2018iw} The dielectric constant in the direction perpendicular to the graphite plane was found to be as low as $\sim 2$ within the layer of water $\sim 7$ \AA\ in thickness. Specifying the projection of the dielectric constant is important for interfacial polarization since the scalar dielectric constant of the bulk transforms into a two-component tensor characterizing the dipolar response perpendicular, $\epsilon_\perp$, and parallel, $\epsilon_\parallel$, to the substrate plane.\cite{Stern:2003cn,Ballenegger:05,Gekle:2012kx,Zhang:2013gq,De-Luca:2016aa} Ref.\ \onlinecite{Fumagalli:2018iw}, therefore, reports $\epsilon_\perp\sim 2$ for water at graphite. A somewhat higher value, $\epsilon_\perp\sim 3.8$, was suggested for water in contact with the negatively charged mica surface,\cite{PhysRevE.64.011605} while $\epsilon_\perp\sim 2$ was suggested for the water-air interface.\cite{Teschke:2005ea}   

In this study, we have addressed the problem of the interfacial dipolar response for a somewhat related type of the interface formed between buckminsterfullerenes C$_{60}^z$ and SPC/E water.\cite{Berendsen:87} Here, $z$ denotes the total charge of the fullerene, which, in practical applications, can be altered by electrochemistry.\cite{Reed:2000aa} We apply molecular dynamics (MD) simulations to study the effects of solute's charge ($-4\le z \le 1$) and water's temperature ($240 \le T\le 360$ K) on the interface susceptibility. The linear dipolar susceptibility of the interface is used to define the interface dielectric constant $\epsilon_\text{int}$, which is a property characterizing the interface and distinct from the bulk dielectric constant $\epsilon$ ($\epsilon\sim 71$ for SPC/E water at 300 K\cite{Fennell:2012ee}). We find that $\epsilon_\text{int}\sim 10-22$ is not significantly affected by temperature, but is much stronger affected by the solute charge. The interface susceptibility  as a function of charge $z$ passes through a spike marking a structural crossover of the hydration shell.

\section{Interface susceptibility} 
The dielectric constant of a bulk dielectric is a material property, independent of sample's shape, because of the locality of the Maxwell field $\mathbf{E}$. The Maxwell field is defined as an ensemble average of the microscopic field $\mathbf{E}_m$, followed by a coarse-graining protocol averaging out molecular-scale oscillations of the microscopic field.\cite{Landau8} The postulated\cite{*} locality of the Maxwell field allows one to relate it to the local polarization density $\mathbf{P}$ through the scalar susceptibility, $\mathbf{P}=\chi\mathbf{E}$. 

Despite this widely adopted reasoning, $\mathbf{E}$ itself is never accessible experimentally and only the line integral $\Delta \phi=\int \mathbf{E}\cdot d\mathbf{l}$ connects the Maxwell field to experimentally accessible voltage difference $\Delta \phi$ between the end points.\cite{EygesBook:72} Dielectric experiments take advantage of the fact that $E_z$ is uniform in a plane capacitor and find the $z$-projection of the Maxwell field $E_z=\Delta \phi/d$ in terms of the separation $d$ between the plates ($z$-axis is perpendicular to capacitor's plates). 

The Maxwell field must be non-uniform in the interface and, by that fact, it becomes a property not accessible to measurements. Also, locality of an inhomogeneous Maxwell field has never been established.  The fundamental difficulty of measuring local fields in non-uniformly polarized dielectrics has long been recognized.\cite{Thompson1872} The only known resolution is to either measure the field produced by a polarized dielectric in vacuum\cite{EygesBook:72} or to measure fields inside small cavities carved within the dielectric.\cite{Thompson1872,DMjcp2:11} The second strategy is realized in experiments recording solvent-induced shifts of optical lines giving access to local cavity fields.\cite{Fried:2015fh}     

The locality of polar response disappears for interfaces, which cannot be characterized by a well-defined scalar susceptibility. In fact, the locality of the response is gone even in the bulk at microscopic length-scales, when bulk dielectric susceptibility $\chi$ is converted to a nonlocal 2-rank tensor response function $\bm{\chi}(\mathbf{r}-\mathbf{r}')$. It reduces to longitudinal and transverse dielectric projections $\chi^{L,T}(k)$ depending on the scalar wavevector $k$ in reciprocal space.\cite{Dolgov:81,Fonseca:90,Raineri:92,Bopp:96} For inhomogeneous polarization encountered in solvation and interfaces, the polar susceptibility, $\bm{\chi}(\mathbf{k}_1,\mathbf{k}_2)$, loses its isotropic symmetry and becomes a function of two wavevectors.\cite{Chandler:93,DMjcp1:04}

A complete microscopic solution for the interface polarization is clearly complex and is only remotely related to dielectric properties of the bulk. One still wonders if a coarse-grained description in terms of an effective susceptibility of the interface can be formulated. It is clear that any such definition will not be unique and is likely to apply to a set of problems for which the response of the interface is well defined. Our focus is on interface polarization in terms of the field it produces within a cavity\cite{Thompson1872,DMjcp2:11} in response to a probe charge. Since the Coulomb law applies also to microscopic fields, this goal can be achieved by a proper reformulation of the boundary value problem. The interface susceptibility giving access to the field inside a void is not necessarily transferrable to other electrostatic problems and, specifically, to another well-established problem where dielectric constant is prominent: the screening of ions in solution. 

It is easy to realize that a length-scale is involved in dielectric screening: the potential of mean force between ions is an oscillatory function of the distance at molecular scale,\cite{Huston:1989is,Rashin:1989aa,Bader:1992hm,Fennell:2009fe} obviously not reducible to a single screening parameter. Likewise, the average polarization density in the interface induced by a spherical ion $\langle P_r\rangle = \langle \mathbf{\hat r}\cdot\mathbf{P}\rangle$ ($\mathbf{\hat r}=\mathbf{r}/r$ is the radial unit vector) is an oscillatory function of the radial distance $r$ as found in simulations by Ballenegger and Hansen\cite{Ballenegger:05} and in a number of follow-up simulation studies of spherical and planar interfaces.\cite{Bonthuis:2012mi,Gekle:2012kx,Schaaf:2015dl,DMjcp3:16,Shi:2016jq} There is obviously a length-scale, dictated by the microscopic structure, specific to the local polarization density of the interface. A definition of a scalar parameter of interfacial polarity based solely on $\mathbf{P}(\mathbf{r})$ is not possible.     

The polarization of the interface $\mathbf{P}(\mathbf{r})$ is not directly accessible experimentally and is not even required if the focus is on the measurable local field inside a solute or cavity in the dielectric. From this perspective, the question at hand is what is the integrated response of the interface to a probe charge. Posed in this way, the problem of an effective susceptibility of the interface can potentially be formulated without a length-scale involved, if a proper coarse-graining formalism is formulated. Cast in this way, the problem becomes somewhat similar to the problem of surface tension, which is clearly a microscopic interfacial property, but without a length-scale involved. One asks the question of what is the integrated response of the interface to altering its surface area. Similarly, we are asking what is the integrated polarization of the interface creating a certain field inside a void. Presenting a formalism to address this question and its application to a realistic interface of charged buckminsterfullerenes in water is the goal of this study.      

The microscopic electric field $\mathbf{E}_m=-\nabla\phi_m$ is expressed in terms of the microscopic electrostatic potential $\phi_m$ in the presence of the external charge density $\rho_0(\mathbf{r})=q\delta(\mathbf{r})$ and the instantaneous density of bound charge $\rho_b=-\nabla\cdot \mathbf{P}$. The fluctuating electrostatic potential satisfies the Laplace equation at each configuration of the liquid
\begin{equation}
\nabla^2 \phi_m  = -4\pi\left[\rho_0 + \rho_b\right] .
\label{eq1}
\end{equation}
The density of bound charge, a scalar field, is in turn expressed through the divergence of the fluctuating polarization field $\mathbf{P}=\mathbf{P}_d-\tfrac{1}{3}\nabla\cdot\mathbf{Q}+\dots$, which includes the dipolar field $\mathbf{P}_d$ and spatial derivatives of densities of higher multipoles,\cite{Jackson:99} starting from the quadrupolar 2-rank tensor $\mathbf{Q}$. When a solute is immersed in a polar liquid, $\rho_b=0$ inside the solute and $\phi_m$ is determined from the standard Laplace equation, $\nabla^2 \phi_m=-4\pi\rho_0$. A significant result here is that one can find a solution for an ensemble-averaged potential $\langle\phi_m\rangle$ provided the boundary conditions are additionally supplied. Therefore, finding the electric field inside a cavity in the dielectric is reduced to the question of formulating proper boundary conditions that preserve, in a coarse-grained manner, some information about the molecular structure of the interface.\cite{DMjcp2:11,DMjcp3:14}

\begin{figure}
	\includegraphics*[clip=true,trim= 0cm 0cm 0cm 0cm,width=5cm]{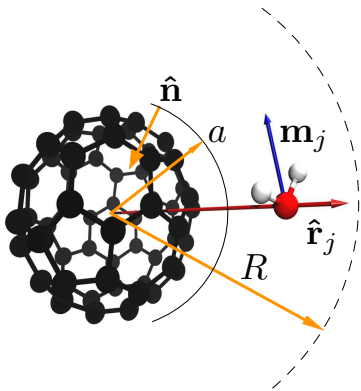}
\caption{Schematics of the C$_{60}$-water interface.  The spherical dividing surface with the radius $a$ (solid line) has the normal unit vector $\mathbf{\widehat n}$ pointing outward from the liquid toward the solute. The spherical region in water with the radius $R$ is used to calculate the total dipole moment of the water molecules within the $R$-sphere, Eq.\ \eqref{eq7}. The scalar normal projection of the dipole moment is calculated by projecting each individual dipole moment $\mathbf{m}_j$ (blue arrow) of the water molecule with $r_j<R$ on the radial unit vector $\mathbf{\widehat r}_j=-\mathbf{\widehat n}_j$, $\mathbf{\widehat r}_j=\mathbf{r}_j/r_j$ (red arrow).  }	
\label{fig:1}
\vskip -0.3cm
\end{figure}

The boundary conditions account for the discontinuity of the electric field at some dividing surface between the solute and the liquid
\begin{equation}
E_{0n} - \langle E_n\rangle = 4\pi\left [\sigma_0 + \langle P_n\rangle\right] .   
\label{eq2} 
\end{equation}
In this equation, $E_{0n}= -\mathbf{\widehat n}\cdot\nabla\phi_0$  is the normal projection of the electric (vacuum) field inside the solute and $\langle E_n\rangle = -\mathbf{\widehat n}\cdot\nabla \langle \phi_m\rangle$ is the normal projection of the ensemble-averaged electric field inside the solvent, both are taken at a nonspecified dividing surface separating the solute from the solvent. The surface normal $\mathbf{\widehat n}$ is directed outward from the solvent into the solute (Fig.\ \ref{fig:1}). Equation \eqref{eq2} averages over the fluctuations of the solute-solvent system by taking ensemble averages $\langle\dots\rangle$. Further, $\sigma_0$ in Eq.\ \eqref{eq2} is the density of free charges at the dividing surface. This component of the electrostatic problem is important for our simulations of charged fullerenes C$_{60}^z$ since the solute charge is distributed over the surface atoms when $z\ne0$.  This charge configuration is different from solvation of small ions for which only $P_n$ enters the boundary condition.\cite{DMjcp3:16} 

The normal projection of the instantaneous polarization density of water $P_n=\mathbf{\widehat n}\cdot \mathbf{P}$ is the main focus of our formalism and of the simulations presented below. Its ensemble average $\sigma_b=\langle P_n\rangle$ on the right-hand side of Eq.\ \eqref{eq2} is the density of the bound (water) charge at the dividing surface.\cite{Landau8} The surface charge density is a parameter quantifying the preferential alignment of dipoles in the interface. Its relation to the bulk properties of the dielectric material is different for liquid and solid dielectrics as can be highlighted by first looking at the results following from the theories of continuum (macroscopic) dielectrics. 

The result for the surface charge density is particularly simple for the spherical symmetry of the dielectric interface and the polarizing electric field. When a probe charge $q$ is placed at the center of a spherical cavity with the radius $a$ carved from a dielectric, the surface charge density becomes $\sigma_b = - (1-\epsilon^{-1})(q/S) = (4\pi)^{-1}(1-\epsilon^{-1})E_{0n}$, where $S=4\pi a^2$ is the surface area.\cite{Jackson:99}  The electrostatic potential of the bound charge inside the dielectric becomes 
\begin{equation}
	\phi_b(r) = - \left(1-\epsilon^{-1}\right) (q/r) .
	\label{eq3}
\end{equation}
The surface charge is, therefore, opposite to the probe charge and screens it. When combined with the vacuum potential $\phi_0(r)=q/r$, Eq.\ \eqref{eq3} yields the standard Coulomb potential screened by the dielectric, $\langle\phi_m(r)\rangle=q/(\epsilon r)$.

Equation \eqref{eq3} assumes that a bulk material property, the dielectric constant here, can define a property of the interface, the dipolar polarization of the interface in our case. This assumption\cite{Maxwell:V1} strictly applies only to solid dielectrics, which can propagate bulk stress through the entire material by means of a uniform strain when the uniform stress/field is applied. One can view the polarization of the interface as preferential alignment of interfacial dipoles uniformly propagating from the bulk. The polarization density in the interface $\mathbf{P}$ is then the same as in the bulk when the dielectric is uniformly polarized (plane capacitor), which is the meaning of the notion of a continuum dielectric described by the boundary conditions of Maxwell's electrostatics.

Liquids do not maintain bulk stress, and the polarization in response to an external field must form in a surface layer of molecular dimension. As mentioned above,  the issues involved are similar to the distinction between the surface tension, a macroscopic property characterizing interface only, and the cohesive energy of the bulk. Drawing from this analogy, bulk dielectric constant does not necessarily describe surface polarization. Two different susceptibilities are, therefore, required: the interface susceptibility to describe polarizability of the interface and bulk dielectric constant. The former describes orientational preferences of the interfacial dipoles, which are strongly affected by the local interfacial structure. The latter describes the buildup of dipolar correlations by chains of mutually induced dipoles\cite{Wertheim:71} producing long-ranged, $\propto r^{-3}$, correlations ultimately responsible for dielectric screening.\cite{Hoye:74} There is no direct link between these two susceptibilities since the alignment of dipoles in the interface is a function of both the liquid and the substrate.        

\begin{figure}
	\includegraphics*[clip=true,trim= 0cm 0.2cm 0cm 0cm,width=7.5cm]{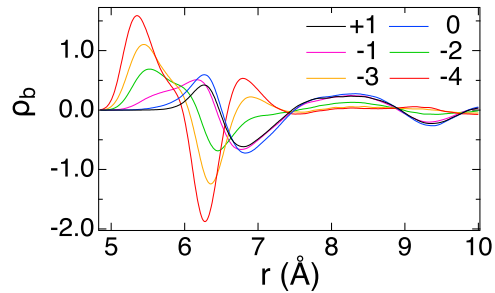}
\caption{Density of bound charge of water $\rho_b(r)$ [Eq.\ \eqref{eq20}] for C$_{60}^z$ with the solute charges indicated in the plot. 
}
\label{fig:2}
\end{figure}

In contrast to the dielectric constant of the bulk, the dipolar susceptibility of the interface has not been uniquely defined. The vector field of the dipole moment density $\mathbf{P}(\mathbf{r})$ is highly oscillatory in the interface and does not provide a scalar susceptibility, even if a specific projection  is taken.\cite{Ballenegger:05,Horvath:2013fe,Shi:2016jq} This is illustrated in Fig.\ \ref{fig:2} where the density of bound charge of water, 
\begin{equation}
  \rho_b(r) = \sum_{i=1}^N\sum_{a=1,3} \langle q_{i,a}\delta (\mathbf{r}-\mathbf{r}_{i,a})\rangle
  \label{eq20}
\end{equation}
 in the interface of buckminsterfullerenes C$_{60}^z$ is shown (the sum here runs over all $i=1,\dots,N$ water molecules with atomic charges $q_{i,a}$ at $\mathbf{r}_{i,a}$, where $a=1-3$ specifies the atoms in the water molecule). 
 
 To avoid uncertainties of arbitrary definitions of the interface susceptibility, we use the susceptibility required for closing the boundary condition in Eq.\ \eqref{eq2} and consequently solve the electrostatic problem inside the void. A linear relation between $E_{0n}$ and $\langle P_n \rangle$ provides such a closure
\begin{equation}
\langle P_n \rangle = \chi_{0n} E_{0n} .
\label{eq4}	
\end{equation}
This route\cite{DMjcp3:14} is applied here to calculate $\chi_{0n}$ from molecular dynamics trajectories and to evaluate the interface dielectric constant $\epsilon_\text{int}\ne \epsilon$ based on the input from numerical simulations.

\begin{figure}
	\includegraphics*[clip=true,trim= 0cm 0.2cm 0cm 0cm,width=7.5cm]{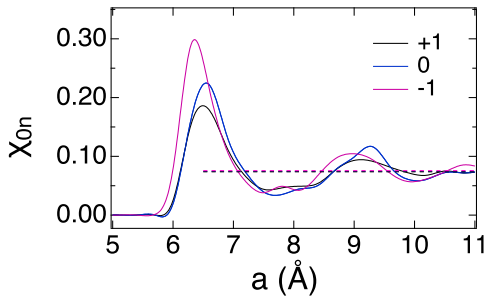}
\caption{Interface susceptibility $\chi_{0n}(a)$ for C$_{60}^z$ with the solute charges indicated in the plot. The dashed lines (nearly indistinguishable on the scale of the plot) show coarse-grained $\chi_{0n}$ calculated from the slopes of $\chi^I$ in Eq.\ \eqref{eq7}. }
\label{fig:3}
\end{figure}

Solutes studied here are neutral and charged fullerenes C$_{60}^z$, with the charge $z$ varied between $z=-4$ and $z=1$. The entire charge $z$ of a charged fullerene is spread over its surface with the charge density $4\pi\sigma_0=ze/a^2= - E_{0n}(a^+)$ (Fig.\ \ref{fig:1}), where $E_{0n}(a^+)$ is the electric field of the solute charges at the outer surface of the fullerene. In contrast, the electric field inside the charged fullerene in zero, $E_{0n}(a^-)=0$, if the surface charge is assumed to be uniformly spread. It is this field, $E_{0n}=E_{0n}(a^-)$ that enters the boundary condition on the left-hand side in Eq.\ \eqref{eq2}.

In the case of the neutral fullerene with $z=0$, we assume that the solute field $E_{0n}$ is produced by the probe charge $q$ placed at the center of the solute. Both cases of charged and neutral fullerenes can be combined in Eq.\ \eqref{eq2} to obtain a closed-form equation\cite{DMjcp3:14} for the dipolar susceptibility $\chi_n$ connecting $\langle P_n\rangle$ to the ensemble-averaged (Maxwell) field $\langle E_n\rangle$:\cite{2016PhRvB..93n4201Z} $\langle P_n\rangle=\chi_n\langle E_n\rangle$. The same expression for the interface dielectric constant $\epsilon_\text{int}-1=4\pi\chi_n$ follows in both cases     
\begin{equation}
\epsilon_\text{int}  = \left[1 - 4\pi \chi_{0n}\right]^{-1} .
\label{eq5}	
\end{equation}

The perturbation theory\cite{DMjcp3:14} gives $\langle P_n\rangle$ as the statistical correlation of the fluctuation $\delta P_n= P_n-\langle P_n\rangle$ with the fluctuation of the solute-solvent Coulomb energy $U^\text{C}$       
\begin{equation}
\langle P_n \rangle = -\beta \langle \delta P_n \delta U^\text{C}\rangle ,
\label{eq6}	
\end{equation}
where $\delta U^\text{C} = U^\text{C}-\langle U^\text{C}\rangle$. Further, $U^\text{C}=Q\phi_s$ is given as the product of the fluctuating electrostatic potential $\phi_s$ of the solvent and the solute charge $Q$: $Q$ is either $q$ (for $z=0$) or $ze$ (for $z\ne0$). By combining Eqs.\ \eqref{eq4} and \eqref{eq6}, the solute charge can be eliminated from the linear  susceptibility 
\begin{equation}
\chi_{0n} = \beta a^2 \langle \delta P_n \delta \phi_s\rangle ,	
\label{eq6-1}
\end{equation}
where $a$ is the radius of the spherical dividing surface and $\delta\phi_s=\phi_s-\langle \phi_s\rangle$. Because of the spherical symmetry of the problem, $\phi_s$ is taken at the center of C$_{60}^z$ for both charged and neutral solutes.

\begin{figure}
	\includegraphics*[clip=true,trim= 0cm 0.2cm 0cm 0cm,width=7.5cm]{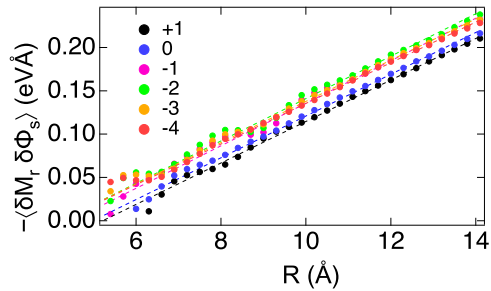}
\caption{Correlation of the radial projection of the shell dipole moment $M_r(R)$ [Eq.\ \eqref{eq8-1}] with the electrostatic potential of the solvent $\phi_s$ at the center of C$_{60}^z$ as a function of the radius $R$ of the spherical shell used to calculate the water dipole moment [Eq.\ \eqref{eq8}]. Calculations are performed at different values of charge $z$ (plots at different temperatures are collected in Figs.\ S5--S8 in supplementary material). The slope of $\chi^\text{I}(R)\propto \langle \delta M_r(R) \delta \phi_s \rangle$ with the radius $R$ determines the interface susceptibility $\chi_{0n}$ [Eq.\ \eqref{eq7}]. The dashed lines are linear fits through the simulation points. }	
\label{fig:4}
\end{figure} 

Equation \eqref{eq6-1} is the integrated form\cite{DMjcp3:16} of the equation given by Ballenegger and Hansen,\cite{Ballenegger:05} in which the two-point correlation function between radial projections of the polarization density $P_r=\mathbf{\hat r}\cdot\mathbf{P}$ needs to be integrated 
\begin{equation}
  \chi_{0n} = 4\pi a^2 \beta \int_a^\infty \langle P_r(a) P_r(r')\rangle dr' .
\end{equation}
The susceptibilities $\chi_{0n}$ in both equations rely on a specific value of the radius $a$ for the dividing surface. This definition is not computationally robust since the dividing surface separating dielectrics, and, even more so, the separation between a liquid and a molecular-scale solute, is not well defined in dielectric theories. This problem is shared not only by solvation theories, where the ``dielectric cavity'' can only be empirically established,\cite{Roux:90,Rick:94} but also by the local polar response discussed in the computational literature.\cite{Stern:2003cn,Ballenegger:05,Bonthuis:2011dq,Gekle:2012kx,Zhang:2013gq,De-Luca:2016aa} Specifically, $\chi_{0n}(a)$ oscillates as a function of the radius $a$ of the dividing sphere onto which both the polarization density and the external field are projected (Fig.\ \ref{fig:3} and Fig.\ S5 in supplementary material). Therefore, in order to arrive at a single robust parameter, averaging oscillations of the polarization density out, a coarse-graining protocol was suggested in Ref.\  \onlinecite{DMjcp3:14}. 

Instead of using $\chi_{0n}(a)$ calculated at a specific $a$ in Eq.\ \eqref{eq6-1}, an average $\chi_{0n}$ over a range of dividing surface radii $a$  it is calculated. Coarse graining of molecular scale interfacial oscillations is performed by calculating the slope of the integrated susceptibility $\chi^\text{I}(R)$ vs the radius $R$ of the spherical region chosen around the solute (Fig.\ \ref{fig:1}) 
\begin{equation}
\chi_{0n} = d\chi^\text{I}/dR 	. 
\label{eq7}
\end{equation}
In turn, the integrated susceptibility  
\begin{equation}
4\pi\chi^\text{I}(R) = - \beta \langle \delta M_r(R) \delta \phi_s \rangle	
\label{eq8}
\end{equation}
is calculated by correlating the fluctuations of the electrostatic potential produced by the solvent at the position of the probe charge  with the radial projection of the total dipole moment of the solvent within the $R$-sphere (Fig.\ \ref{fig:1}) 
\begin{equation}
M_r(R) = \sum_{r_j < R} \mathbf{m}_j \cdot \mathbf{\widehat r}_j .	
\label{eq8-1}
\end{equation}
Here, $\mathbf{m}_j$ are the dipole moments of the water molecules within the $R$-sphere. They are projected on their corresponding radial unit vectors $\mathbf{\widehat r}_j=\mathbf{r}_j/r_j$. By constructing the dipole moment according to Eq.\ \eqref{eq8-1}, we have neglected the contribution of the quadrupolar polarization density to $\mathbf{P}$. Molecular quadrupole contributes to the surface potential,\cite{Wilson:1989kx} but is expected to disappear when the integral of $\nabla\cdot\mathbf{Q}$ is taken over the closed volume between the dividing surface and the $R$-sphere (Fig.\ \ref{fig:1}).   

The application of this formalism to the calculation of the interface dielectric constant $\epsilon_\text{int}$ is illustrated in Fig.\ \ref{fig:4}, where $\phi_s$ is calculated at the geometrical center of the solute. It shows the correlation $-\langle \delta M_r(R) \delta \phi_s  \rangle$ [Eq.\ \eqref{eq8}] as a function of the radius $R$ of the spherical region chosen to calculate $M_r(R)$. Oscillations at low $R$ reflect the molecular structure of water in the interface. The coarse-grained susceptibility $\chi_{0n}$ effectively averages out the oscillations of $\chi_{0n}(a)$ in the interface. This is indicated  by the horizontal dashed lines in Fig.\ \ref{fig:3} comparing $\chi_{0n}$ with $\chi_{0n}(a)$.  The interface susceptibilities follow from the slopes of the plots in Fig.\ \ref{fig:4} according to Eq.\ \eqref{eq7}. In what follows, we discuss the results of this analysis applied to simulations of fullerenes C$_{60}^z$ in SPC/E\cite{Berendsen:87} water.    

\section{Results} 
The geometries and charge distributions of C$_{60}^z$ were calculated with the density functional theory as described in supplementary material. These charges were combined with CHARMM22/OPLS parameters for carbon atoms and used as the force field for classical MD simulations performed with NAMD 2.9 software program.\cite{Phillips:2005qv} The nonuniform distribution of atomic charge in charged fullerenes is caused by Jahn-Teller distortions of icosahedral symmetry characteristic of the neutral fullerene\cite{Niklas:2018kw} (see supplementary material for discussion). However, we found that DFT charges and the uniform distribution of atomic charge $z/60$ produce indistinguishable results for the interfacial structure of hydration water (Fig.\ S1 in supplementary material). Therefore, in addition to integer charges $z=+1,\dots,-4$ from DFT, uniformly distributed charges $z/60$,  $z=-0.25,-0.5,-1.5,-1.0,-1.75,-2.5$ were employed in simulations at $T=300$ K. This set of simulations allowed us to obtain the dependence of the interfacial dipolar susceptibility on the solute charge and to connect it to the structural crossover of hydration water occurring with increasing $|z|$ (see below). 

The C$_{60}^z$ solutes were hydrated with 2413 SPC/E water\cite{Berendsen:87} molecules. Initial equilibration with NPT was followed by NVT simulations at different temperatures (240, 260, 280, 300, 320, 340 and 360 K). A typical simulation length was 110 ns. A detailed list of simulation times is provided in Table S1 in supplementary material, along with other details of the simulation protocol. 

\begin{figure}
\includegraphics*[clip=true,trim= 0cm 0cm 0cm 0cm,width=7.5cm]{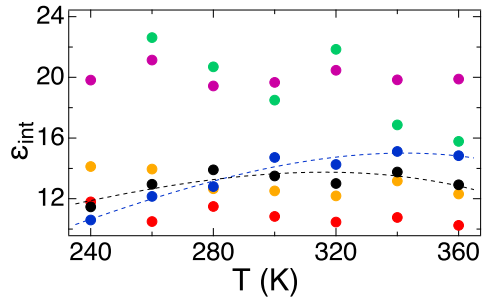}
\caption{Interface dielectric constant vs $T$ for different charges $z$ of C$_{60}^z$: $z=+1$ (black), $z=0$ (blue), $z=-1$ (magenta), $z=-2$ (green), $z=-3$ (orange), $z=-4$ (red). The electrostatic potential of water $\phi_s$ entering Eq.\ \eqref{eq8} is calculated at the center of C$_{60}^z$; the dashed lines are polynomial fits through the points for $z=0$ (blue) and $z=1$ (black). }	
\label{fig:5} 
\vskip -0.3cm
\end{figure}

The main finding of our simulations and their analysis is a significant reduction of $\epsilon_\text{int}$ compared to the bulk dielectric constant ($\epsilon\sim  71$ for SPC/E\cite{Fennell:2012ee}). This result is relevant for the electrostatic boundary value problem for which the bulk value $\epsilon$ is often used. In contrast, our formulation suggests that the interface dielectric constant  $\epsilon_\text{int}$, carrying molecular properties of the interface, should be used in place of $\epsilon$ in the boundary conditions (Eq.\ \eqref{eq2}) for the Laplace equation. The resulting values of $\epsilon_\text{int}$ depending on temperature and solute charge are summarized in Fig.\ \ref{fig:5} and in Table S2 in supplementary material. There are some noticeable changes with temperature, particularly for the neutral fullerene (a dashed blue line in Fig.\ \ref{fig:5}). However, a much stronger variation of the interface dielectric constant is observed with the solute charge $z$, reflecting a structural crossover in the first hydration shell for charged solutes.    

\begin{figure}
\includegraphics*[clip=true,trim= 0cm 0cm 0cm 0cm,width=7.5cm]{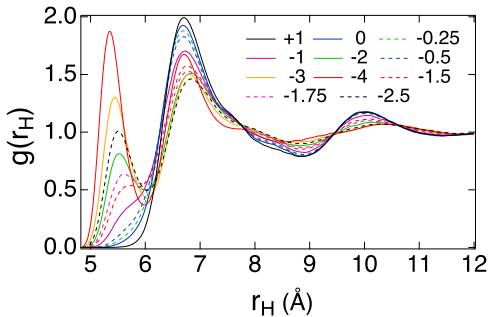}
\caption{Radial distribution function for the hydrogens of SPC/E water at different charges $z$ of C$_{60}^z$ indicated in the plot ($T=300$ K). }	
\label{fig:6} 
\vskip -0.3cm
\end{figure}

The structure of water interfacing charged fullerenes is significantly altered compared to the bulk. Water's density in the first hydration layer increases with increasing charge. More importantly, interfacial water undergoes a structural transition from preferential orientations specific to hydrophobic interfaces to an orientational structure characteristic of charged substrates.\cite{DMpccp2:18}  Signatures of the structural transition are seen already at $z=-1,-2$, and the new structure of the hydration shell is fully formed at $z=-3,-4$ (Fig.\ \ref{fig:6}). The transition is accompanied by the destruction of the hydrogen-bond network of the hydration shell and the release of dangling O-H bonds pointing to fullerene's center by nearly every water molecule out of $\sim 40$  first-shell waters at $z=-3,-4$. 

The change in preferential orientations strongly affects the distribution of the bound charge in the interface: the appearance of the positive and negative peaks corresponding to dangling O-H groups is clearly seen in the interfacial density of bound charge shown in Fig.\ \ref{fig:2}. It is also seen in the growing peak of the solute-hydrogen pair distribution function, which is separated from the second peak by the O-H bond length $\sim 1$ \AA\ (Fig.\ \ref{fig:6}).  This type of crossover from an in-plane orientation of interfacial waters, typical for hydrophobic solvation,\cite{Lee:84,Shen:2006gf} to a large population of dangling O-H bonds was recorded by x-ray absorption of water on gold substates under a negative bias.\cite{VelascoVelez:2014ea}

\begin{figure}
\includegraphics*[clip=true,trim= 0cm 0cm 0cm 0cm,width=7.5cm]{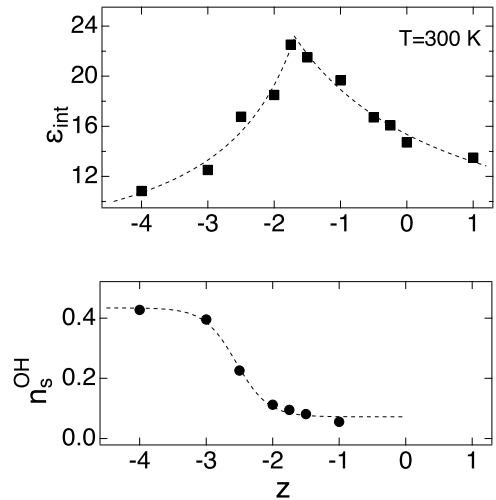}
\caption{Upper panel: Interface dielectric constant at $T=300$ K vs the fullerene charge $z$. The dashed lines are fits to Curie-type functions $a+b/(z-z0)$. Lower panel: The average number of dangling O-H bonds in the first hydration layer of C$_{60}^z$. The dashed line is a fit to a hyperbolic tangent function often appearing in mean-field theories of phase transitions.\cite{Stanley:87} }	
\label{fig:7} 
\vskip -0.3cm
\end{figure}

The release of dangling O-H bonds\cite{Shen:2006gf,Davis:2013if,VelascoVelez:2014ea} at the point of crossover can be characterized by the order parameter given by the fraction $n_s^\text{OH}$ of dangling O-H in the first hydration shell.\cite{DMpccp2:18} It is calculated from the relative area of the closest peak in the solute-hydrogen pair distribution function (Fig.\ \ref{fig:6}). This order parameter is plotted in the lower panel of Fig.\ \ref{fig:7} vs the solute charge $z$ (also see Table S3). The dashed line fitting through the points is a hyperbolic tangent function often appearing in mean-field theories of phase transitions.\cite{Stanley:87} The structural crossover, carrying some phenomenology of bulk phase transitions, is accompanied by a spike in the interface dielectric constant shown in the upper panel of Fig.\ \ref{fig:7}. The dashed lines fitting the points are of Curie type: $a+b/(z-z_0)$. This functionality appears in the Landau theory of phase transitions\cite{Stanley:87} when the quadratic term in the free energy functional is taken in the form $\propto (z-z_0)$ (the standard Curie law follows from $\propto (T-T_0)$). One, however, should not anticipate a Curie-type singularity found for susceptibilities characterizing bulk phase transitions because of a small number of water molecules involved. Nevertheless, the phenomenology of Curie's law is approximately retained for the interface dielectric constant approaching the crossover point. 

Our simulations employ a non-polarizable water model accounting only for the response of the nuclear coordinates. As a minimum, electronic susceptibility $\epsilon_\infty-1$ should be added to $\epsilon_\text{int}$ to account for electronic polarizability of water molecules. Here, $\epsilon_\infty$ is the high-frequency dielectric constant\cite{Boettcher:73}. Its precise value for water is not known: values between squared refractive index, $\epsilon_\infty\sim 1.8$, and $\epsilon_\infty\sim 4.2$ have been proposed in the literature.\cite{Kaatze:1997gg}  The situation is potentially more complex, as is seen for the water-air interface. The average, mean-field dipole of water changes in the water-air interface from a bulk value, enhanced relative to the gas phase, to the gas-phase dipole.\cite{Cipcigan:2015cq} While the interface with a solute or with a planar substrate is obviously distinct from the water-air interface, a nonuniform interfacial electric field might lead to an effective water dipole different from the bulk.    

The interface dielectric constant considered here is a gauge of the integral ability of the interface to polarize in response to a probe charge placed inside a void in a polar liquid. In contrast, thermally-driven fluctuations of the overall shell dipole $\mathbf{M}(R)$ can be gauged by either its variance $\langle [\delta\mathbf{M}(R)]^2\rangle$\cite{Marti:2006kd,Mukherjee:2017ix} or by the scalar product, $\langle\delta\mathbf{M}(R)\cdot \delta\mathbf{M}_s\rangle$, with the total dipole moment of the sample $\mathbf{M}_s$.\cite{Stern:2003cn,DMcpl:11} By this measure, dipole fluctuations actually increase, and not decrease, compared to the bulk for hydrophilic solutes.\cite{Marti:2006kd,DMcpl:11} The variance of the shell dipoles scales with the local density\cite{Marti:2006kd,DMcpl:11} and, consistent with this logic, a recent paper\cite{Sato:2018gb} reports a drop of water's dielectric constant at model hydrophobic interfaces with reduced surface density (dewetting). However, the Kirkwood formula was incorrectly applied\cite{Sato:2018gb} to the calculations of the dielectric tensor in the interface. The Kirkwood formula is derived by tracing the dipolar susceptibility over its longitudinal and transverse components.\cite{SPH:81,Madden:84} For the slab geometry, those correspond to linear responses perpendicular and parallel to the slab.\cite{DMjcp2:16,DMNlinDiel} The Kirkwood equation, therefore, cannot be applied to components of the dielectric tensor, which can carry different symmetries in respect to the longitudinal and transverse dipolar susceptibilities.\cite{Madden:84} 

The variance of the shell dipole moment, which can be used to characterize dipolar fluctuations in the interface, does not directly enter the electrostatic boundary-value problem. It is only the normal projection of the dipole moment that defines the interface susceptibility\cite{DMjcp3:14,DMjcp3:16} and is required for electrostatics. Only this susceptibility is lower in the interface than in the bulk. This result implies  suppression of the interfacial response in the normal direction. Dipoles in the interface, frustrated by the local fields and geometric constraints,\cite{Lee:84,Zhang:2013gq,VelascoVelez:2014ea,Wen:2016df} do not develop the complete dielectric screening of the bulk material. The deviation between $\epsilon_\text{int}$ and $\epsilon$ is less pronounced for the fluid of dipolar hard spheres interfacing a repulsive void.\cite{DMjcp3:14}  The origin of a large difference between $\epsilon_\text{int}$ and $\epsilon$ for water is likely a signature of its specific interfacial orientational structure, which is difficult to characterize in more detail without expanding the interface susceptibility in basis functions sensitive to orientational dipolar order.\cite{Besford:2018db}

\section{Conclusions}
The computational formalism used here allows direct access to interface susceptibility from configurations produced by computer simulations. The required property is the cross-correlation of the radial projection of the dipole moment of the solvation shell with the electrostatic potential of the solvent inside the solute.

This computational formalism has been applied to simulations of a realistic interface chemically similar to the water-graphite interface studied experimentally in Ref.\ \onlinecite{Fumagalli:2018iw}, where dielectric constant $\epsilon_\perp\sim 2$ was reported for thin, $\sim 7$ \AA, films of water.  Values of the interface dielectric constant $\epsilon_\text{int}\sim 10-22$ are reported here for charged fullerenes interfacing SPC/E water. Interface dielectric constants in the same range, $\sim 6-9$, were previously calculated from simulations of model Lennard-Jones solutes of different sizes in TIP3P water.\cite{DMjcp3:16} The interface dielectric constant $\sim 11-15$ ($T=240-360$ K) for the neutral fullerene is consistent with $\epsilon_\text{int}\sim 9$ calculated for the smallest LJ solute with the first peak of the solute-oxygen radial distribution function at $\sim 5.5$ \AA\ (compared to $\sim 6.75$ \AA\ for C$_{60}^0$). Since the non-polarizable force fields for water miss the response of the electronic polarizability, our results likely constitute the lower bound for the interface susceptibility. The interface dielectric constant is the property of an interfacial layer of water of molecular scale and is physically distinct from the bulk dielectric constant reflecting dipolar correlations in the bulk ($\epsilon\sim 71$ for SPC/E water\cite{Fennell:2012ee}).

Increasing the charge $|z|$ of hydrated fullerenes leads to a structural transition of hydration water. It is characterized by breaking the interfacial network of hydrogen bonds and the release of dangling O-H bonds pointing toward the solute. The interface dielectric constant marks this structural crossover with a spike.

\section*{Supplementary material}
See supplementary material for the simulation protocols, data analysis, and additional plots of the interface susceptibility at different temperatures.

\acknowledgments 
This research was supported by the National Science Foundation (CHE-1800243). This work used the Extreme Science and Engineering Discovery Environment (XSEDE) through allocation TG-MCB080071. 


%

\end{document}